\documentclass[structabstract]{aa}  
\usepackage{graphicx}

\begin{document}
   \title{Low and high velocity clouds produced by young stellar clusters}
%
%
   \author{A. Rodr\'\i guez-Gonz\'alez
          \inst{1},
          A. C. Raga\inst{1} \and J. Cant\'o\inst{2}
          }

   \institute{Instituto de Ciencias Nucleares, 
Universidad Nacional Aut\'onoma de M\'exico, 
Ap. 70-543, M\'exico, D.F., CP: 04510, M\'exico,
              \email{ary, raga@nucleares.unam.mx}
         \and
          Instituto de Astronom\'{\i}a, 
Universidad Nacional Aut\'onoma de M\'exico,
Ap. 70-468, M\'exico, D.F., CP:04510, M\'exico
             }

   \date{}

 
 \abstract
{Intermediate and high velocity HI clouds rain onto the plane
of our Galaxy. They are observed at heights of between 500 and 1500~pc,
falling onto the Galactic plane at velocities from 50 to 140~km~s$^{-1}$.}
  {To explain the origin of these clouds, we present a galactic fountain model, driven
by the wind from a super stellar cluster (SSC).}
  {We solve the equations for a steady, radiative de Laval nozzle flow.
We consider two effects not considered previously
in astrophysical nozzle flow models: cooling functions for different
metallicities, and the direct action of the galactic gravitational
field on the gas flowing along the nozzle.}
  {For an adiabatic nozzle flow, the gravity acting
directly on the gas within the nozzle ``stalls'' the nozzle
flow for initial wind velocities lower than the escape velocity
from the Galaxy. For the same wind velocity, a radiative
nozzle flow stalls at lower altitudes above the galactic plane.
We find that SSC winds with velocities of $v_w=500 - 800$~km~s$^{-1}$
produce nozzles stall at heights of $x_m=1 - 15$~kpc. The stalled
nozzle flow then rains back onto the galactic plane
at velocities in the range observed in intermediate and high
velocity HI clouds.}
{We study a nozzle flow driven by a wind from
a SSC close to the Galactic
centre. We find that for velocities within the range expected
for a SSC wind, we can produce nozzle flows that stall above
the galactic plane. These stalled flows  produce
 cool, infalling clouds with velocities similar to those
 of intermediate and high velocity HI clouds.}

\keywords{Hydrodynamics - ISM: jets and outflows - Galaxies: star clusters -
Galaxies: starburst - Galaxies: intergalactic medium}
\titlerunning{IVCs and HVCs produced by young stellar clusters}
\maketitle

\section{Introduction}

The observations of Munch \& Zirin (1961) showed absorption (mostly
Ca~II) lines of interstellar clouds at distances of 500-1500 pc above
the Galactic plane. These high altitude interstellar clouds have 
been described by many authors as neutral hydrogen (HI) clouds at
distances of between 300 pc and 12 kpc with infall velocities between
$-50$ and $-140$~km~s$^{-1}$.

The clouds have been divided into two categories:
intermediate-velocity clouds (IVCs) and high-velocity clouds (HVCs).
IVCs have radial velocities in the range of 30 km s$^{-1}$ 
$\leq$ $\vert$v$_{LSR}$$\vert$ $\leq$  90 km s$^{-1}$
 with respect to the LSR, and  HVCs have radial velocities
$\vert$v$_{LSR}$$\vert$ $\geq$ 90 km s$^{-1}$. In the
Galaxy (and in other galaxies), the IVCs have typical altitudes 
of between 300 pc and 2.5 kpc, and  masses of up to $\sim 10^4$ M$_\odot$.
HVCs have been observed at  higher altitudes (of up to $\sim 12$~kpc).
Several complexes of HVCs with masses of $\sim 10^5 - 10^6$ M$_\odot$
have been observed (see, e.~g., Wakker \& van Woerden 1997).
The general characteristics of HI clouds in
the  Galaxy were discussed  by Dickey \& Lockman (1990) and
Wakker \& van Woerden (1997). Their general chemical properties were
 described by Lu et al. (1998), Wakker et al. (1999), Murphy et
al. (2000), Sembach et al. (2000), and  Bluhm et al. (2001).

Significant progress has been made in exploring the 
distribution of IVCs  and HVCs in the halo of the Galaxy
(see, e.g., Dettmar 2005). However, the origin of these clouds is 
 unclear. Oort (1970) proposed 
that high altitude clouds (in the Galaxy) are formed in past episodes of
star formation. Several authors have used this scenario to explain the
origin of the IVCs. However, the HVCs have been considered to be
 extragalactic objects that are about to merge with the Galaxy 
(Fraternali \& Binney 2006, Melioli et al. 2008). An alternative 
scenario was  proposed by Blitz et al. (1999), who suggested that 
HVC and IVC could be relics of the early stages of Galactic formation.

It has also been proposed that both IVCs and HVCs are formed in the
disk of our  galaxy, in the so-called ``galactic fountains''. These
models were described by Shapiro \& Field (1976), Houck \&
Bregman (1990),  and  de Gouveia Dal Pino et al. (2009).  
In this scenario, the cloud formation is fed by chimneys of
hot, ascending gas that cools and falls back onto the Galactic
disk in the form of discrete clouds. Observations 
indicate that chimneys are indeed generated in other
galaxies by multiple supernovae (SNe)
explosions (Veilleux, Cecil \& Bland-Hawthorn 2005, Konstantopoulos 
et al. 2008).

De Gouveia Dal Pino et al. (2009)
showed that SNe explosions can lead to the formation of
hot superbubbles that drive a supersonic wind out of the 
galactic disk. From 3D, radiative hydrodynamical simulations,
 they found that the galactic fountains
driven by SNe may reach altitudes of up to $\sim 5$~kpc
in the halo and thus allow the formation of  IVCs. However,
the HVCs (found at higher altitudes)
require a different production mechanism.

We propose that HVCs (and possibly also some of the IVCs) could
be formed by fountains driven by super stellar clusters (SSCs).
SSCs are dense clusters of 
young massive stars. These stellar clusters contain hundreds or
thousands of very young, energetic stars, and have stellar densities
far greater than those seen in normal OB associations. These star clusters
have ages of $\sim 1 - 10$ Myr, radii of $\sim$ 1 -10 pc, total cluster
masses of $10^3$ - $10^6$ M$_\odot$  (Melo et al. 2005
reported a  mean  mass per stellar cluster of $\sim 2 \times  10^5$
$M_\odot$,  for M82), and typical cluster wind velocities of
$500 - 1500$ km s$^{-1}$ range (see also, Silich et al. 2004). The
central stellar densities of SSCs have values of
up to $\sim$ $10^5$ M$_\odot$ pc$^{-3}$. However, 
SSCs with older ages and/or higher masses do exist
(Walcher et al. 2006 reports  a cluster with 6 $\times$ 10$^7$ M$_\odot$). 
On the other hand, the  
metallicity of the gas reinserted (via supernova explosions and stellar 
winds) into the stellar cluster medium, reaches supersolar values 
(1 $\leq$ Z $\leq$ 14, see also Meynet \& Maeder 2002 and 
Tenorio-Tagle, Silich, and Mu\~noz-Tu\~n\'on 2003) during most
of the cluster evolution.

SSCs have been observed in a wide range of star-forming galaxies, 
such as merging systems (NGC 4038/4039, Whitmore  \& Schweizer 1995), 
dwarf galaxies (Henize  2-10, Johnson et al.  2000), classical 
starbursts (M82, Gallagher  \& Smith 1999, Melo et al.  2005), 
as well as in the Galaxy
(Arches Cluster: Nagata et al.  1995; Cotera et al. 1996; Serabyn,
Shupe  \& Figer 1998) amongst many  other systems (for a review, see
Whitmore 2000).

In this paper, we present a model for the formation of
low and high velocity HI clouds by means of radiative nozzle flows driven
by the wind from a SSC. The nozzle flow is produced by  the interaction
of the shocked SSC wind with the stratified halo and the gravitational
potential of the Galaxy. In our model, we assume that the SSC that 
produces to the fountain is close to the Galactic centre.
This is an appropriate assumption if we consider the wind
from the Arches cluster or from other possible massive (but
still undetected) stellar clusters close to the Galactic centre.
In galaxies such
as M82 and NGC253, it appears that we do see outflows ejected from
SSC's in the central regions of these galaxies (see
Rodr\'\i guez-Gonz\'alez et al. 2008).

The paper is organized as follows.
In Section 2, we describe the equations for a radiative nozzle flow,
and in Section 3 we discuss the analytic solution obtained
in the adiabatic case. In Section 4, we describe the Galactic
potential and hot gas distribution, which are used
in Section 5 to compute numerical integrations of the radiative nozzle flow 
equations. Finally, in Section 6 we discuss the application to these
models for low and high velocity HI cloud formation, and we present
our conclusions in Section 7.
\section{Equations for a radiative nozzle flow}
We assume that the interaction of a wind from a SSC (located
close to the centre of the galaxy) with the
surrounding interstellar medium (ISM) leads to the formation 
of an approximately
spherical shock. The thermalized, post-shock material is then
channeled into two oppositely directed nozzles because
of the pressure stratification of the environmental material.
 In the context of the generation of jets by active galactic nuclei, 
this scenario was first explored by Blandford \& Rees (1974)

We consider a one-dimensional, stationary, radiative nozzle flow
under the action of a gravitational potential. We first write the
energy equation in the ``entropy conservation'' form
\begin{equation}
\label{ec:3}
\frac{d}{dx}P \rho^{-\gamma}=-\frac{(\gamma-1)L}{\rho^\gamma u}\,.
\end{equation}
where $x$ is the direction of propagation of the flow, $P$ is
the gas pressure, $\rho$ the density, $u$ the flow velocity,
$\gamma$ the specific heat ratio, and $L$ is the energy loss
per unit time and volume.
We now consider the momentum equation in the form
\begin{equation}
\label{ec:mom}
\frac{1}{\rho}\frac{d P}{dx}=-\frac{d}{dx}\left[\frac{u^2}{2}+\Phi(x)\right]\,,
\end{equation}
\noindent
where $\Phi(x)$ is the gravitational potential, which we take to be
a known function. Equations (\ref{ec:3}) and (\ref{ec:mom})
 can be combined to obtain
\begin{equation}
\label{ec:bern}
\frac{d}{dx}\left[\frac{u^2}{2}+\frac{c_s^2}{\gamma-1}+
\Phi(x) \right]=-\frac{L}{\rho u}\,,
\end{equation}
with
\begin{equation}
\label{ec:sound}
c^2_s=\frac{\gamma P}{\rho}\,.
\end{equation}
Equation (\ref{ec:bern}) simplifies to Bernoulli's theorem for
the case of a non-radiative flow (i. e., for $L=0$).

Following Blandford \& Rees (1974), we now assume that the
nozzle flow has a slowly varying cross-section, so that
the pressure of the flow always adjusts to a value close
to the pressure $P_e(x)$ of the surrounding environment.
If one also knows the cooling function $L$ as a function of
the density and  temperature of the flow,
the nozzle flow can then be obtained by integrating the system
described by eqs. (\ref{ec:3}) and (\ref{ec:bern}).

The initial conditions for this integration are obtained by assuming
that the initially isotropic wind from the cluster (of mass-loss
rate ${\dot M}$ and terminal velocity $v_w$) follows a strong,
spherical shock before flowing along the nozzle. Assuming that the
shock is strong, we have a post-shock velocity
\begin{equation}
\label{ec:vps}
v_{ps}={{\gamma-1}\over {\gamma+1}}v_w\,,
\end{equation}
and a post-shock sound speed
\begin{equation}
\label{ec:cps}
c_{ps}={\sqrt{2\gamma(\gamma-1)}\over {\gamma+1}} v_w\,.
\end{equation}
Following Raga \& Cant\'o (1989), we assume that the initial velocity
and sound speed along the nozzle flow are $v_0\approx v_{ps}$ and
$c_0\approx c_{ps}$, respectively (see Eqs. \ref{ec:vps} and
\ref{ec:cps}). The initial density $\rho_0=\rho(x=0)$ of the nozzle
flow can be obtained from the $P(x)=P_e(x)$ lateral
pressure-balance condition (see above)
\begin{equation}
\label{ec:rho0}
\rho_0={\gamma P_e(0)\over c_{ps}^2}\,,
\end{equation}
where $P_e(0)$ is the (known) environmental pressure at $x=0$, and
$c_{ps}$ is given by Eq. (\ref{ec:cps}). The initial conditions
for the integration of the nozzle flow problem (Eqs.
\ref{ec:3} and \ref{ec:bern}) are then given by Eqs.
(\ref{ec:vps}-\ref{ec:rho0}).

Finally, the mass conservation equation for a flow of
slowly varying cross-section
\begin{equation}
\label{ec:mass}
{\dot M}=\pi R^2 \rho u\,,
\end{equation}
where ${\dot M}$ is the mass-loss rate,
can be used to calculate the cylindrical radius
$R$ of the cross-section as a function of position $x$
along the nozzle flow.

\section{The adiabatic solution}

For a non-radiative flow (i.e., with $L=0$),
Eq. (\ref{ec:bern}) can be trivially integrated
to obtain
\begin{equation}
\label{ec:bernad}
{u^2\over 2}+{c_s^2\over {\gamma-1}}+\Phi(x)={v_w^2\over 2}+\Phi_0\,,
\end{equation}
where $\Phi_0=\Phi(x=0)$, and the constant on the right is obtained by
setting $u=v_{ps}$ and $c_s=c_{ps}$ (see Eqs. \ref{ec:vps}
and \ref{ec:cps}) at $x=0$.

For $L=0$, from Eqs. (\ref{ec:bernad}), (\ref{ec:mom}),
(\ref{ec:sound}) and (\ref{ec:cps}) and the lateral pressure balance
condition, we obtain the solution:
\begin{equation}
\label{ec:usol}
u^2(x)=v_w^2\left\{1-{4\gamma\over {(\gamma+1)^2}}
\left[{P_e(x)\over P_0}\right]^{(\gamma-1)/\gamma}\right\}
+2\left[\Phi_0-\Phi(x)\right]\,,
\end{equation}
where $P_e(x)$ is the environmental pressure stratification,
and $\Phi(x)$ is the gravitational potential. If these two functions
are known, Eq. (\ref{ec:usol}) directly gives the velocity
along the nozzle flow as a function of position $x$. We should
note that this solution is equivalent to the one of Blandford
\& Rees (1974), but also includes the direct effect of the gravitational
force on the material within the nozzle flow.

It is easy to check that for $x=0$, Eq. (\ref{ec:usol})
correctly gives $u(0)=c_{ps}$ (see Eq. \ref{ec:cps}), the
imposed initial value for the flow velocity. For $x\to \infty$, for
all finite mass environmental stratifications we have $P_e(x)\to 0$
and $\Phi(x)\to 0$ (setting the zero of the potential at infinity).
We then obtain
\begin{equation}
\label{ec:uinfty}
u(x)\to u_\infty=\sqrt{v_w^2+2\Phi_0}\,,\,\,\,\,{\rm for}\,\, x\to \infty\,.
\end{equation}
Because $\Phi_0<0$, this equation clearly has a
physical solution only if
\begin{equation}
\label{ec:vesc}
v_w>v_{esc}=\sqrt{-2\Phi_0}\,,
\end{equation}
in other words, the nozzle flow extends out to infinity only if
the velocity of the wind (which is producing the nozzle flow) exceeds
the escape velocity.
\begin{figure}
\centering
\includegraphics[width=\columnwidth]{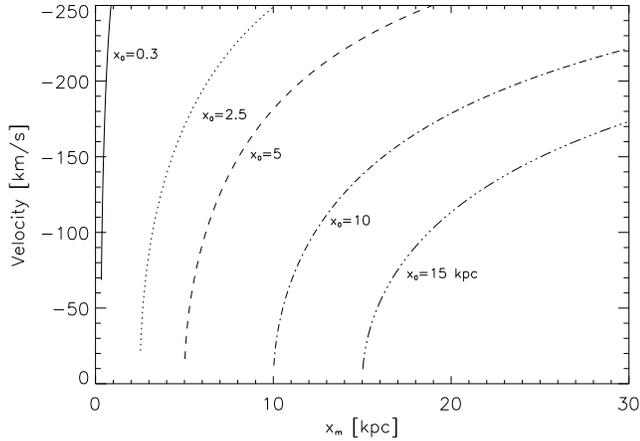}
\caption{Free-fall velocity as a function of $x_m$ for different final
heights $x_0=0.3$, 2.5, 5, 10 and 15 kpc (solid, dotted, dashed, dash-dot, and
dash-dot-dot lines, respectively).}
\end{figure}
If this condition is not met, the nozzle flow reaches a maximum
height $x_m$ (which can be computed from Eq. \ref{ec:usol}
setting $u=0$), at which it stalls. However, since the escape velocity
from the centre of our galaxy has a value
$v_{esc}\approx 700$~km~s$^{-1}$, a wind
from a cluster of massive stars (with $v_w\sim 1000$~km~s$^{-1}$)
satisfies the condition shown in Eq.(\ref{ec:vesc}) , and therefore
 produces an adiabatic nozzle flow that escapes from the Galaxy. In
 the following sections, we show that this is not necessarily the case
 for radiative nozzle flows.

\section{The Galactic potential and hot gas distribution}

We assume that the halo has a uniform temperature. An isothermal halo is
an assumption that is consistent with the observed diffuse,
soft X-ray background emision (Garmire et al. 1992). The Galactic gas
density $\rho_g$ and thermal pressure $P_g$ are given by
(Espresate et al. 2002)
\begin{equation}
\label{ec:deng}
\rho_g(x)=\rho_{g,0}\exp^{[\Phi_0-\Phi(x)]/c^2_g},
\end{equation}
and,
\begin{equation}
\label{ec:pe}
P_g(x)=P_{0}\exp^{[\Phi_0-\Phi(x)]/c^2_g},
\end{equation}
where, $P_0=\rho_{g,0}c^2_g$ and $\rho_{g,0}$ are the pressure and density, 
(respectively), of the hot gas at the Galactic centre, and $c_g$ is the 
(uniform) isothermal sound speed of the halo.
For the gravitational potential $\Phi(x)$, we use the model proposed by 
Allen \& Santill\'an (1991) and Espresate et al. (2002). These
authors defined three components for the Galactic mass: (1) a 
{\it nucleus}, represented by a spherically symmetric distribution, 
(2) a {\it disk component}, (3) a {\it spherical, massive halo}. Hence,
 the gravitational potential is given by the sum of these three contributions.

The units are such that if $x$ is in kiloparsecs and
$M_1$, $M_2$, and $M_3$ are in Galactic mass units (gmu=2.32$\times$10$^7$ 
M$_\odot$), the units for $\Phi(x)$ are 100
km$^2$  s$^{-2}$ (Espresate 2002). The equations for the potentials
along the rotation axis of the Galaxy, for the different components
are as follows\\
1. the nucleus is represented by
\begin{equation}
\label{ec:pot1}
\Phi_1(x)=\frac{M_1}{(x^2+b^2_1)^{1/2}},
\end{equation}
where, $M_1$=606 gmu and $b_1=0.3873$ kpc;\\
2. the disk component is represented by
\begin{equation}
\label{ec:pot2}
\Phi_2(x)=\frac{M_2}{a_2+(x^2+b^2_2)^{1/2}},
\end{equation}
where, $M_2=3690.0$ gmu, $a_2=5.3178$ kpc, and $b_2=0.25$ kpc; and,\\
3. finally, a halo truncated at 100 kpc (see also Allen \&
Santill\'an  1991), has a potential
\begin{equation}
\label{ec:pot3}
\Phi_3(x)=b_3\left(1+\ln\left[1+\left(\frac{x}{a_3}\right)^{1.02}\right]\right)
-c_3,
\end{equation}
where, $a_3=12.0$ kpc, $b_3=377.04$ gmu/kpc, and $c_3=1201$ gmu/kpc. The total
gravitational potential at any point $x$ along the rotation axis
of the galaxy is then given by
\begin{equation}
\label{phit}
\Phi(x)=\Phi_1(x)+\Phi_2(x)+\Phi_3(x).
\end{equation}
This expression for the gravitational potential (from Espresate et
al. 2002) implies a total mass for the Galaxy of $M_T= 9.0 \times 10^{11}$
M$_\odot$, and a {\it local} circular velocity of 220 km s$^{-1}$ 
and $r_\odot=8.5$ kpc.
With this potential, we can compute the free-fall velocity of
material injected at a height $x_m$ above the galactic plane (by
a nozzle flow originating close to the galactic centre). By the
time that the material injected (with a velocity $\approx 0$) at
$x_m$ has fallen  to a height $x_0$ above the galactic
plane, it has acquired a velocity:
\begin{equation}
\label{eq:vf}
v_f^2=\Phi(x_m)-\Phi(x_0)\,.
\end{equation}

Figure 1 shows $v_f$ as a function of $x_m$ for different final
heights $x_0$, for values of $x_0$ that are typical
of the observed altitudes of IVCs and HVCs. The resulting values of
$v_f$ are used in the following sections to determine the conditions
that have to be satisfied by nozzle flows  to produce clouds with
velocities similar to those of IVCs and HVCs.
\section{The models}

We consider a nozzle flow fed by a wind from a massive stellar
cluster situated close to the Galactic centre. The wind travels through
a shock and then forms a nozzle aligned with the rotation axis of the
Galaxy, which will is to the gravitational potential given
by Eq. (\ref{phit}), and the environmental pressure stratification
given by Eqs. (\ref{ec:deng}-\ref{ec:pe}).

We fix the environmental pressure stratification and the gravitational
potential as described in the previous section by assuming
a temperature of $10^6$~K and a density on the galactic
plane of 0.5~cm$^{-3}$ for the hot halo. The remaining free parameters
of the model are then the cluster wind velocity $v_w$, its mass-loss rate
${\dot M}$, and metallicity $Z$. The initial flow velocity, sound
speed, and density of the flow are functions of only the wind velocity $v_w$
and the pressure of the hot environmental gas close to the galactic
plane (see Eqs. \ref{ec:vps}-\ref{ec:rho0}). The metallicity
$Z$ is important for determining the cooling rate.
The mass-loss rate is only important
for determining the radius of the nozzle flow cross-section  
(see Eq. \ref{ec:mass}), so that we choose a single mass-loss 
rate $\dot{M}$=$10^{-2}$ M$_\odot$ yr$^{-1}$ for all models.

This is a reasonable mass-loss rate in the following sense. 
To reach a height of $\sim 10$~kpc above the galactic plane, the
nozzle flow must exist for a time $\sim 10$~kpc$/v_w\sim 10$~Myr
(where $v_w\sim 1000$~km~s$^{-1}$ approximate
cluster wind velocity). This timescale coincides with the lifetime
of the stellar population of a SSC (Leitherer \& Heckman
1995). During this lifetime, a cluster of $\sim 10000$ massive
stars (with a total mass of $\sim 10^6$ M$_\odot$) will have lost
$\sim 10$\%\ of its initial mass. If we consider a much smaller
cluster such as, e.g., the Arches cluster near the Galactic centre
(which has only $\sim 100$ massive stars), it would be necessary
to consider a much lower  mass-loss rate ($\sim 10^{-4}$ M$_\odot$
yr$^{-1}$). However, the mass-loss rate from the cluster enters only
in the calculation of the cylindrical radius of the nozzle flow
(see Eq. 8 and Fig. 2), and the dynamical characteristics
of the flow do not depend on the value of the mass loss.

We should note that a cluster of $10^4 - 10^5$ massive stars will feed a $\sim
10^5 - 10^6$ M$_\odot$ nozzle flow, which in principle could produce
the mass observed in an HVC complex (see Wakker \& van Woerden
1997). A cluster with 100-1000 massive stars will
produce a $\sim 10^3 - 10^4$ M$_\odot$ nozzle flow, corresponding
to the typical mass of an IVC (see de Gouveia Dal Pino et al. 2009).

We now carry out numerical integrations of Eqs. (\ref{ec:3})
and (\ref{ec:bern}) with the initial conditions given by
Eqs. (\ref{ec:vps}-\ref{ec:rho0}) for different values
of $v_w$ and $Z$. We use the cooling function of Raymond, Cox \& Smith (1976),
who tabulated the cooling rate as a function of the temperature
$T$ and metallicity $Z$ of the gas (a density dependence $\propto \rho^2$,
appropriate to a low-density regime cooling, is assumed).
\begin{figure}
\centering
\includegraphics[width=\columnwidth]{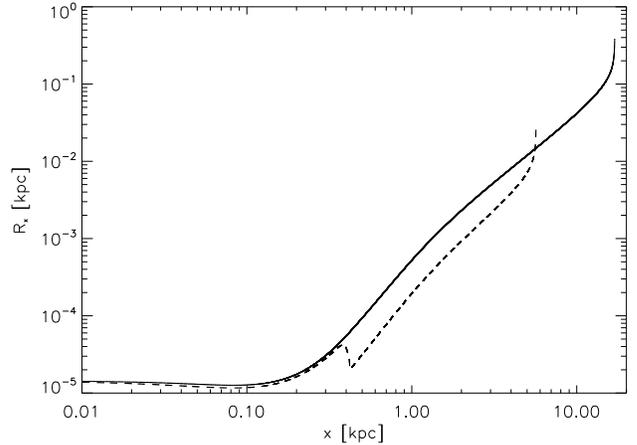}
\caption{The radius of the nozzle flow cross-section as a function 
of height $x$ above the galactic plane.  The solid line shows the $Z=1$ model
 and the dashed line shows $Z=5\, Z_\odot$ model.}
\end{figure}
Figures 2-5 show the results obtained for models with $v_w=700$~km~s$^{-1}$
and $Z=1$, and 5 solar metallicities. Figure 2 shows the radius
of the nozzle flow cross-section as a function of height $x$ above
the galactic plane. The $Z=1$ model (solid line) shows a radius
that decreases for $x<0.10$~kpc, and then increases monotonically
at greater heights. This behaviour is similar to that found
in adiabatic nozzles, in which the nozzle radius decreases
in the region of subsonic flow, and increases with distance
after the sonic transition. The radius of the nozzle flow
diverges at $x_m\approx 15$~kpc. This divergence was not found
in previous nozzle solutions (see, e.g., Blandford \& Rees
1974 and Raga \& Cant\'o 1989), because they did not include
the effect of the gravitational force acting directly on the
material within the nozzle flow.
From Figure 2, we see that the solution for $Z=5$ (with
stronger radiative cooling) shows a sudden collapse
of the nozzle radius at $x\approx 40$~kpc. This collapse
is caused by the strong radiative cooling at temperatures below
$10^6$~K, and is qualitatively similar to the results found
for cooling nozzle flows by Raga \& Cant\'o (1989).
\begin{figure}
\centering
\includegraphics[width=\columnwidth]{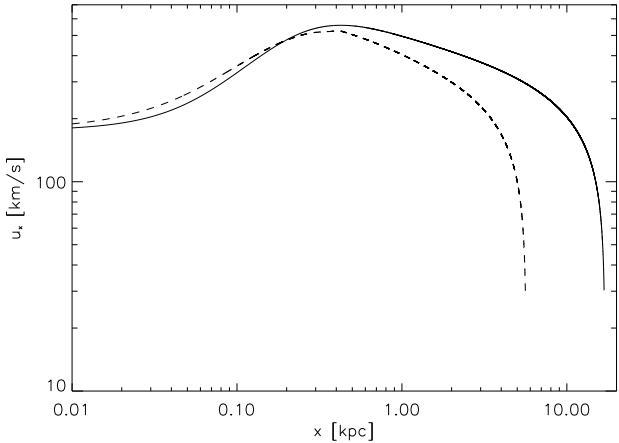}
\caption{The gas velocity as a function of height $x$ above the galactic plane.
The lines have the same description as Fig. 2. }
\end{figure}
Figure 3 shows the velocity along the nozzle flow for the
two models. The flow velocity first grows with increasing $x$,
reaches a maximum, and then collapses to zero at the point
where the nozzle radius diverges (see figure 2).
\begin{figure}
\centering
\includegraphics[width=\columnwidth]{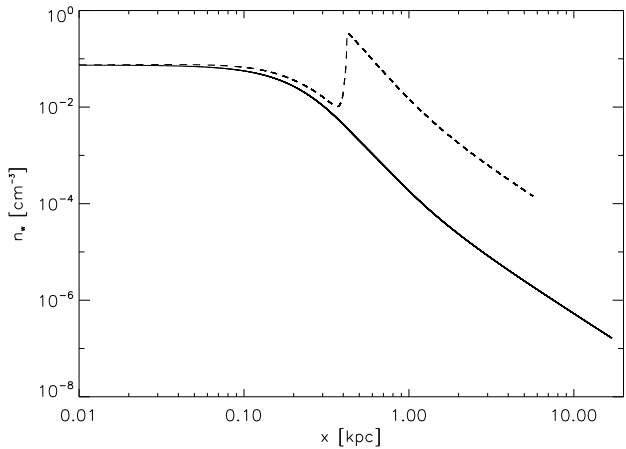}
\caption{The numerical gas density as a function of height $x$ above the
 galactic plane. The lines have the same description of Fig.2. }
\end{figure}
We show in Figure 4 the density as a function of position along 
the nozzle flow. For $Z=1$, a monotonically decresing
density vs. $x$ relationship is found. For $Z=5$, a sharp
increase in the density is found at the position in which
the nozzle radius collapses (see Fig. 2).

Finally, in Figure 5 we show the temperature structure
of the nozzle flow. The $Z=5$ model shows a sudden collapse
at $x\approx 40$~kpc, which is absent in the $Z=1$ model.
\begin{figure}
\centering
\includegraphics[width=\columnwidth]{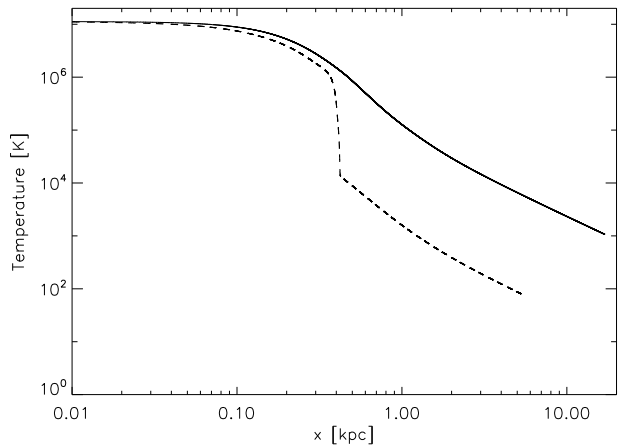}
\caption{The gas temperature as a function of height $x$ above the 
galactic plane. The lines have the same description as  Fig. 2. }
\end{figure}
To explore of parameter space, we have run models with
$v_w=500 - 800$~km~s$^{-1}$ (in $\Delta v_w=50$~km~s$^{-1}$ steps)
and $Z=1 - 10$ (in $\Delta Z=1$ steps). From these models,
we computed the maximum height $x_m$ reached by the
nozzle flow. The values obtained for $x_m$ as a function of
$v_w$ and $Z$ are shown in Figure 6.

From this figure, we see that flows with higher wind velocity
reach larger heights $x_m$. Higher values of $x_m$
are also obtained for decreasing metallicities. The radiative nozzle
flows inject cool material at $x_m$, which will then fall back
onto the galactic plane in the form of dense, neutral clumps.
When these clumps have reached a height $x_0$, they will
have the infall velocities given in Figure 1.

\section{Discussion: low and high velocity HI clouds}

For the adiabatic case, we obtain nozzle flow solutions that
extend out to infinity for cluster wind velicities of
$v_w>v_{esc}$, where $v_{esc}$ is the escape velocity 
from the position of the stellar cluster that feeds the flow
(see Section 3). For $v_w<v_{esc}$, the flow stalls at a finite
distance $x_m$ from the galaxy. As the nozzle stalls, its cylindrical
radius diverges, producing a layer of neutrally buoyant gas
at a distance $\sim x_m$ from the wind source.
If this gas cools radiatively, it will become negatively
buoyant and eventually fall back onto the galactic plane,
reaching velocities similar to the free-fall velocity (see
Fig. 1).

We then integrated the equations numerically for a
radiative nozzle flow. We  included the cooling function
of Raymond, Cox \& Smith (1976) tabulated as a function of temperature for
a set of different metallicities (assuming a ``low density
regime'' $\propto \rho^2$ density dependence). We assume
that the nozzle flow propagates away from the central region
of our galaxy, within an isothermal halo subject to the
Galactic potential of Espresate et al. (2002) and Allen
\& Santill\'an (1991).

We have run a grid of radiative nozzle models for cluster winds with
velocities $500\leq v_w\leq 800$~km~s$^{-1}$ and
metallicities $1<Z<10\,Z_\odot$. From these models,
we found that the radiative cooling lowers the height
$x_m$ at which the nozzle flow stalls. Therefore,
for higher material  metallicities within
the nozzle flows (resulting in higher cooling rates),
we obtained lower values of $x_m$ (see Fig. 6).

In the models with $Z=1\,Z_\odot$,
as the cluster wind velocity increases from
$v_w=500$ to 750 km~s$^{-1}$, the value
of $x_m$ increases from 1 to 50~kpc. For higher
values of $v_w$, the value of $x_m$ rapidly diverges,
indicating that the nozzle flows do not stall (and
therefore never produce negatively buoyant material
falling back onto the galaxy).

In the models with $Z=5\,Z_\odot$, the same
range of $x_m$ values (i.e., from 1 to 50~kpc)
is obtained for cluster wind velocities ranging
from $v_w=620$ to 770~km~s$^{-1}$. For $v_w$ approaching
1000~km~s$^{-1}$, the nozzle flows never stall, regardless
of the metallicity.

Therefore, from our models we could obtain nozzle flows
that stall at heights ranging from 1 to 50~kpc for cluster
wind velocities ranging from $\sim 500$ to 800~km~s$^{-1}$
(and solar abundances $Z=1 - 10$).

Let us now consider low velocity clouds, which are observed at heights 
$x_0 \sim 0.3 - 2.5$~kpc and have infall velocities $< 100$~km~s$^{-1}$. 
From Fig. 1, we see that these clouds could only have originated in
initial heights, $x_m<4$. From Fig. 6. we see that this height
corresponds to the stalling distances of nozzles with
cluster wind velocities $v_w< 620$~km~s$^{-1}$ for
$Z=1\,Z_\odot$ and $v_w< 720$~km~s$^{-1}$ for
$Z=10\,Z_\odot$.
\begin{figure}
\centering
\includegraphics[width=1.06\columnwidth]{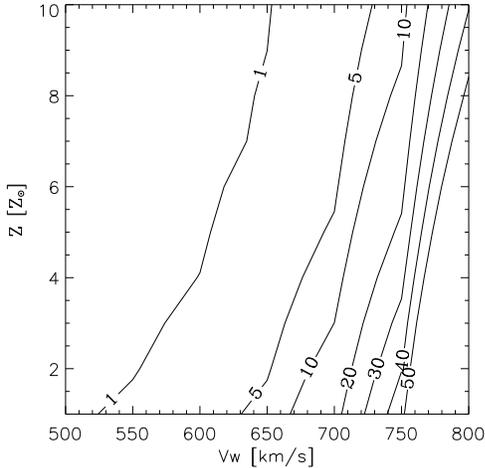}
\caption{Isocontours of maximum height $x_m$ (in kpc) reached 
by the nozzle flow as function of the metallicity Z and cluster wind
$V_w$.}
\end{figure}
High velocity clouds are observed at heights $x_0 >12$~kpc
and have infall velocities $100\to 200$~km~s$^{-1}$. From
Fig. 1, we see that clouds with infall velocities of
150~km~s$^{-1}$ at $x_0=15$~kpc (these being representative
values for high velocity clouds) have originated
at initial heights $x_m\approx 26$~kpc. From Fig. 6,
we see that this height corresponds to the stalling heights
of nozzles with $v_w=720$~km~s$^{-1}$ (for $Z=1\,Z_\odot$)
up to $v_w=780$~km~s$^{-1}$ (for $Z=10\,Z_\odot$).

In this way, we see that a sequence of nozzle models of increasing
$v_w$ with $500<v_w<800$~km~s$^{-1}$ naturally explains the range
of infall velocities observed from low to high velocity HI clouds.
Nozzle flows produced by cluster winds with higher $v_w$ do not
stall, and therefore will not produce large amounts of negatively
buoyant material at high altitudes that could form the observed
HI clouds.

\section{Conclusions}

 We have presented a de Laval nozzle model for galactic
fountain flows. In this flow, the wind emitted (in our case)
by a stellar cluster is thermalized in a wind shock, and
is then reaccelerated out of the galactic plane into a collimated
flow as a result of the pressure stratifiction of the
 surrounding environment.

The equations for this kind of flow were first
studied in the context of the production of jet flows from
AGN by Blandford \& Rees (1974), and in the context of jets
from young stars by K\"onigl (1982). Raga \& Cant\'o (1989) studied
the effect of radiative cooling within the nozzle flow (also
in the context of jets from young stars).

We have written the equations for a nozzle flow, including both
a radiative cooling term and the action of the gravitational
force on the material within the nozzle flow. This latter effect
is caused because the gravitational force (in our case,
because of the Galactic potential) producing the environmental
stratification also acts directly on the material flowing
along the nozzle. This effect has not been included in any
of  previously published astrophysical nozzle flow models.

From numerical integrations of the radiative nozzle flow equations,
we found that cluster winds with $600<v_w<750$~km~s$^{-1}$ produce
nozzles that stall at heights $3<x_m<50$~kpc above the galactic
centre (see Fig. 6). The negatively buoyant clumps that become
separated from the stalled flows fall back towards the galactic plane
at velocities $<200$~km~s$^{-1}$ (see Fig. 1). This velocity range
covers the infall velocities found for both low and high velocity
HI clouds.

From this study, we conclude that the high and low velocity clouds
observed to be falling onto the Galactic disk could indeed be
produced by winds from massive stellar clusters close to the Galactic
centre. The condition for these clusters to produce the observed
clouds is that the cluster wind velocity $v_w$ should be lower
than $\sim 750$~km~s$^{-1}$, with a somewhat higher or lower
velocity limit depending on the metallicity $Z$ of the gas in
the cluster winds (see Fig. 6). Clusters such as the Arches 
(Stevens \& Hartwell 2003) and the Quintuplet clusters (Rockefeller et al. 
2005) are clear candidates for driving this type of nozzle flow.

The Arches and Quintuplet clusters eject masses of the order
of the mass in an IVC (see Section 5). Therefore, several clusters
such as these are necessary to produce
an appreciable fraction of all observed IVC's. A population
of this clusters might be present in the Galactic centre
(with several still undetected clusters). On the other hand, a complex of 
HVC's has a combined mass of $\sim 10^5 - 10^6$ M$_\odot$ (see Section 1, 5). 
To eject these masses from the Galaxy, we would need considerably more massive
SSC's (with $\sim 10^4 - 10^5$ massive stars). It appears unlikely that
if these clusters were present around the Galactic centre they would
not have yet been detected. Therefore, it appears more likely
that these clusters (if they exist) are located far away from the
Galactic centre. It would therefore be interesting to explore
models of nozzle flows produced by clusters
at some distance from the Galactic centre.

We should mention the work of W\"unsch et al. (2008). These
authors showed that the presence of thermal instabilites inside
the wind from a very massive SSC (with more than $8 \times 10^6$ solar
masses) results in a substantial decrease in the total mass-loss
rate of the cluster wind. This effect would serve to moderate
the strength of fountain flows produced by these clusters.

To proceed with this study, it will be necessary to
remove at least some of the simplifying assumptions of our
present model. For example, we have assumed that the nozzle
flow has a uniform velocity across the nozzle cross-section.
The removal of this assumption would probably lead to flows in which
an outer, slower envelope could stall, while a higher velocity
core (moving more repidly than the escape velocity) could continue flowing
out to large distances from the Galaxy. In this a model, cluster
winds with  higher values of $v_w$ might be able to
produce nozzles that form negatively buoyant, infalling clumps.

Inn our work, we have also assumed that the cluster producing
 the nozzle flow is located close to the Galactic centre. It
would be interesting to study off-centre
cluster winds, which should produce nozzle flows with curved
paths, because HVC's could indeed be formed by massive clusters at some
distance from the Galactic centre (see above).

Finally, it would also be interesting to study the
spatial distribution of the infalling clumps (produced by a nozzle flow)
on the galactic plane. To address these issues, it will
probably be necessary to carry out full, 3D gasdynamic simulations
similar to those of de Gouveia Dal Pino et al. (2009, who considered
the  production of high and low velocity clouds
by SN explosions).

\begin{acknowledgements}
We acknowledge support from the CONACyT grant 61547. We thank an
anonymous referee for helpful comments. ACR acknowledges Steve Shore
for interesting discussions ($\sim 20$ years ago) on the subject of
buoyancy in nozzle flows.
\end{acknowledgements}

%
%
%

%
\end{document}